\documentclass[conference]{IEEEtran}
\usepackage[letterpaper, top=0.7in, bottom=1in, left=0.58in, right=0.58in]{geometry}

\usepackage{blindtext, graphicx}

\usepackage{authblk}
\usepackage{balance}  
\usepackage{fixltx2e}
\usepackage{algorithmicx}
\usepackage{algorithm}
\usepackage[noend]{algpseudocode}
\usepackage{url}

\usepackage{enumitem}
\usepackage{multirow}
\usepackage{graphicx}
\usepackage{balance}  
\usepackage{url}
\usepackage[acronym,toc,shortcuts]{glossaries}
\usepackage{epsfig}
\usepackage{amssymb}
\usepackage{amsthm}
\usepackage{multirow}
\usepackage{caption}
\usepackage{subcaption}
\usepackage{siunitx}
\usepackage{soul}
\usepackage{algorithm}
\usepackage{algpseudocode}
\usepackage{csquotes}
\usepackage{bbm}
\usepackage{dsfont}
\usepackage{enumitem}
\usepackage{adjustbox}
\usepackage{array}
\usepackage{multirow}
\usepackage{fancyhdr}
\usepackage[T1]{fontenc}
\usepackage{balance}
\usepackage[table]{xcolor}
\usepackage{comment}

\usepackage{longtable}

\usepackage[acronym,toc,shortcuts]{glossaries}

\graphicspath{{figures/}}
\usepackage{blindtext, graphicx}
\usepackage{soul}

\usepackage[acronym,toc,shortcuts]{glossaries}
\usepackage{amssymb}

\usepackage{longtable}
\usepackage{makecell, boldline}
\usepackage{hhline}
\usepackage{enumitem}
\usepackage{adjustbox}
\usepackage{array}
\usepackage{graphicx}
\usepackage{multirow}
\usepackage{fancyhdr}
\usepackage[T1]{fontenc}
\usepackage{balance}
\usepackage{listings}
\usepackage{url}
\usepackage{xcolor, colortbl}
\usepackage{booktabs}
\usepackage{comment}

\graphicspath{{figures/}}
\newcommand{\mycomment}[1]{}

\begin{document}

\title{Analysis of Post-Quantum Cryptography in User Equipment in 5G and Beyond}

\author{Sanzida Hoque$^{\dagger}$, Abdullah Aydeger$^{\dagger}$,  Engin~Zeydan$^{\ast}$, and Madhusanka Liyanage$^{\varnothing}$ \\ $^{\dagger}$ Dept. of  Electrical Engineering and Computer Science, Florida Institute of Technology, Melbourne, FL, USA \\ $^{\ast}$Centre Tecnològic de Telecomunicacions de Catalunya (CTTC), Barcelona, Spain, 08860. \\ $^{\varnothing}$Network Softwarization and Security Labs (NetsLab), University College Dublin, Ireland\\
\protect Email: \{shoque2023, aaydeger\}@fit.edu, engin.zeydan@cttc.cat, madhusanka@ucd.ie }

\maketitle

\begin{abstract}
The advent of quantum computing threatens the security of classical public-key cryptographic systems, prompting the transition to post-quantum cryptography (PQC). While PQC has been analyzed in theory, its performance in practical wireless communication environments remains underexplored. This paper presents a detailed implementation and performance evaluation of NIST-selected PQC algorithms in user equipment (UE) to UE communications over 5G networks. Using a full 5G emulation stack (Open5GS and UERANSIM) and PQC-enabled TLS 1.3 via BoringSSL and liboqs, we examine key encapsulation mechanisms and digital signature schemes across realistic network conditions. We evaluate performance based on handshake latency, CPU and memory usage, bandwidth, and retransmission rates, under varying cryptographic configurations and client loads. Our findings show that ML-KEM with ML-DSA offers the best efficiency for latency-sensitive applications, while SPHINCS+ and HQC combinations incur higher computational and transmission overheads, making them unsuitable for security-critical but time-sensitive 5G scenarios. 
\end{abstract}


\section{Introduction}

The emergence of quantum computers threatens to undermine the security foundations of modern digital communication. Quantum algorithms, in particular the Shor algorithm, are capable of efficiently solving the hard mathematical problems, integer factorization and discrete logarithms, that underlie widely deployed public key cryptographic systems such as RSA, Digital Signature Algorithm (DSA), and elliptic curve cryptography (ECC) \cite{shor1999polynomial, bernstein2017post}. This imminent threat has accelerated global efforts to transition to quantum-resistant cryptographic algorithms. In response, the U.S. National Institute of Standards and Technology (NIST) initiated the Post-Quantum Cryptography (PQC) Standardization Project, which culminated in the selection of algorithms such as ML-KEM and HQC for key encapsulation and ML-DSA, FALCON and SPHINCS+ for digital signatures \cite{bos2018crystals, nist-pqc}. These schemes are based on difficult problems, such as module lattices and hash-based constructions, which are assumed to be secure against quantum adversaries.

As industry and academia begin to integrate these algorithms into real-world systems, investigating their practicality in constrained and performance-sensitive environments becomes critical. Fifth-generation (5G) wireless networks, which promise improved mobile broadband, ultra-reliable low-latency communication (URLLC), and massive machine-type communication (mMTC), are one area where this challenge is particularly acute. The cryptographic requirements for User Equipment (UE)-to-UE connections are stringent. Limited device computing capacity, tight latency constraints,and constrained wireless bandwidths pose practical deployment challenges for PQC \cite{arfaoui2018security}.  While preliminary evaluations of PQC performance on general processors and embedded platforms have been conducted \cite{hulsing2023sphincs, kannwischer2022improving, liu2017high}, there is still a big gap in understanding their behavior in the context of wireless mobile communication. In particular, the overhead caused by key generation, encapsulation/decapsulation, and digital signature under realistic 5G radio conditions has not yet been comprehensively investigated. Furthermore, the integration of PQC into application layer security protocols, such as TLS 1.3, in UE-to-UE channels raises critical questions about latency, throughput, and resource consumption under typical and heavy-load wireless environments.

In this paper, our contribution includes proposing and discussing PQC algorithm adaptations within TLS for UE-to-UE communications, and presenting a detailed experimental evaluation of NIST-approved PQC algorithms that incorporate key encapsulation mechanisms (KEMs) and digital signature schemes in the context of 5G UE-to-UE communications. We implement selected PQC primitives on 5G emulated Virtual Machines (VM) and evaluate them in realistic radio and network scenarios, focusing on metrics such as handshake latency, computational load, and end-to-end throughput. By bridging the gap between cryptographic security and wireless system performance, this work aims to inform the future integration of quantum-resilient cryptography into wireless standards and ensure that 5G and beyond can remain secure in the post-quantum era.
\begingroup
\renewcommand\thefootnote{}\footnotetext{\textcolor{blue}{This paper has been accepted as a regular paper at LCN 2025 and will appear in the conference proceedings. \textcopyright\ 2025 IEEE. The final version will be published by IEEE and the copyright will belong to IEEE.}}%
\addtocounter{footnote}{-1}
\endgroup

\section{Related Work}

The integration of post-quantum cryptography into mobile networks has attracted a lot of attention, especially in the context of 5G systems \cite{hoque2024exploring, zhou2023survey}. Some early studies have focused on evaluating the performance of PQC algorithms on mobile and embedded platforms. Saarinen conducted an analysis of the energy consumption of various candidates for PQC in Cortex-M4-based systems, highlighting the trade-offs between computational efficiency and communication overhead \cite{saarinen2020mobile}. In the area of 5G authentication protocols, Ko et al. \cite{ko20255g} proposed 5G-AKA-HPQC, a hybrid authentication method that combines classical elliptic curve cryptography with PQC mechanisms. Their work has shown that the integration of PQC into existing 5G authentication frameworks is feasible and that they achieve forward secrecy and quantum resilience without significant performance degradation. Joudah et al. \cite{joudah2024implementing} proposed a key generation and key exchange mechanism using Kyber (aka ML-KEM) to improve the security and performance of the 5G-AKA protocol. Hanna et al. \cite{hanna2024integrating} proposed an end-to-end TLS framework for 5G communications, integrating Falcon512, Dilithium2 (aka ML-DSA65), and SPHINCS+ to mitigate quantum attacks while maintaining compatibility through a VPN-based tunneling approach. Experimental findings indicated that Falcon512 attained the optimal balance between security and low latency, closely paralleling classical TLS performance. Another research \cite{scalise2024applied} incorporates PQC KEMs into the free5GC core network to enhance the security of VNF communications over TLS v1.3, exhibiting negligible latency effects and little data overhead during UE connections. The results demonstrate that Kyber (ML-KEM) KEMs surpass traditional X25519 in connection duration assessments, underscoring the promise of improved quantum-safe key exchange without sacrificing performance. Further research by Demir et al. \cite{demir2025performance} analyzed the performance of NIST-selected PQC algorithms such as Kyber (ML-KEM) and Dilithium (ML-DSA) in telecommunication environments. Their results suggest that these algorithms could be efficiently implemented in 5G networks, although challenges related to infrastructure upgrades and interoperability remain. 

The practical deployment of PQC in mobile networks has also been investigated by the industry. For example, Thales and SK Telecom \cite{thales} conducted a pilot project to implement PQC-based encryption for 5G SIM cards using the CRYSTALS-Kyber (ML-KEM) algorithm to protect subscriber identities from quantum threats. Similarly, Apple announced the integration of PQ3, a hybrid post-quantum cryptographic protocol, into its iMessage platform to protect users' communications from future quantum attacks \cite{wired}. 

Despite these advances, there is a gap in the literature regarding the experimental evaluation of PQC algorithms specifically for UE-to-UE communication in 5G networks. While existing studies have addressed the integration of PQC at the network and application layers, the particular challenges of direct device-to-device communication, such as latency constraints and limited computational resources, have not yet been thoroughly investigated. This paper aims to fill this gap by providing an empirical analysis of NIST-approved PQC algorithms in the context of 5G UE-to-UE communications, focusing on performance metrics that are critical for real-world deployment. 

\section{System Design}
In this section, we provide our system architecture and operational workflow, and detail our proposed PQC-based approach step-by-step into the 5G network.

\begin{figure}
    \centering
    \includegraphics[width=0.7\linewidth]{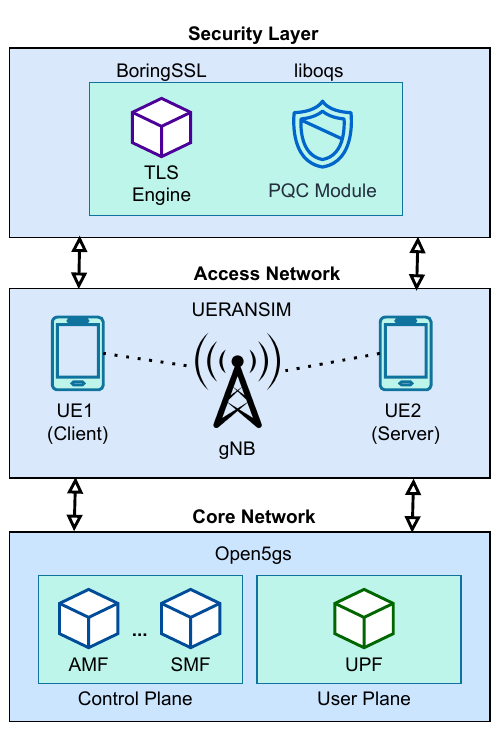}
    \caption{System Architecture }
    \label{fig:sysArch}
    \vspace{-0.1in}
\end{figure}

\subsection{System Architecture}
The system architecture for post-quantum secure UE-to-UE communication has been designed to simulate a comprehensive 5G network environment.
As shown in Fig. \ref{fig:sysArch}, the architecture is divided into three main components: the core network, the access network, and the security layer.

{\textit{Core Network:}} The core network supports key functions such as the Access and Mobility Management Function (AMF), Session Management Function (SMF), User Plane Function (UPF), Authentication Server Function (AUSF), and Network Repository Function (NRF). It includes the control plane (CP) \cite{guttman2018path} and the user plane (UP)\cite{hsieh2021design}, comprising these Network Functions. The AMF manages UE registration and mobility management and ensures seamless communication between UEs. The SMF manages the establishment, modification, and termination of data sessions and enables dynamic allocation of data paths. The UPF serves as a data gateway that handles packet forwarding and routing, which is essential for measuring throughput and latency in UE-to-UE communication.

{\textit{Access Network:}} Access Network consists of UEs and gNBs (gNodeB) that facilitate the generation of network traffic and the establishment of data connections between UEs. 


{\textit{Security Layer:}} The security layer in the system architecture is responsible for establishing and maintaining secure communication channels between UEs using cryptographic protocols.  The security layer works in UEs and handles important cryptographic operations during the TLS handshake and data exchange phases. 




\subsection{Operational Workflow}

This post-quantum secure UE-to-UE communication analysis workflow is divided into four main phases: Initialization, TLS Handshake Process, Data Exchange, and Performance Analysis.

\textbf{\textit{Initialization:}} The process begins with the configuration of the virtual machines (VMs) intended for each network component. The core network components (CP and UP), the access network components (gNB and UEs), and the security layer (BoringSSL, liboqs) are instantiated. The client and server UEs are configured to initiate and respond to PQC-enabled TLS handshakes. The server is configured to support multiple PQC algorithm combinations, e.g. hqc and mlkem for key exchange and falcon for digital signatures.

\textbf{\textit{TLS Handshake Process:}}
The handshake process is systematically recorded in the experiment logs, which document each step in establishing a secure session as displayed in Fig. \ref{fig:seq}. The steps are explained below:

\begin{figure}
    \centering
    \includegraphics[width=1\linewidth]{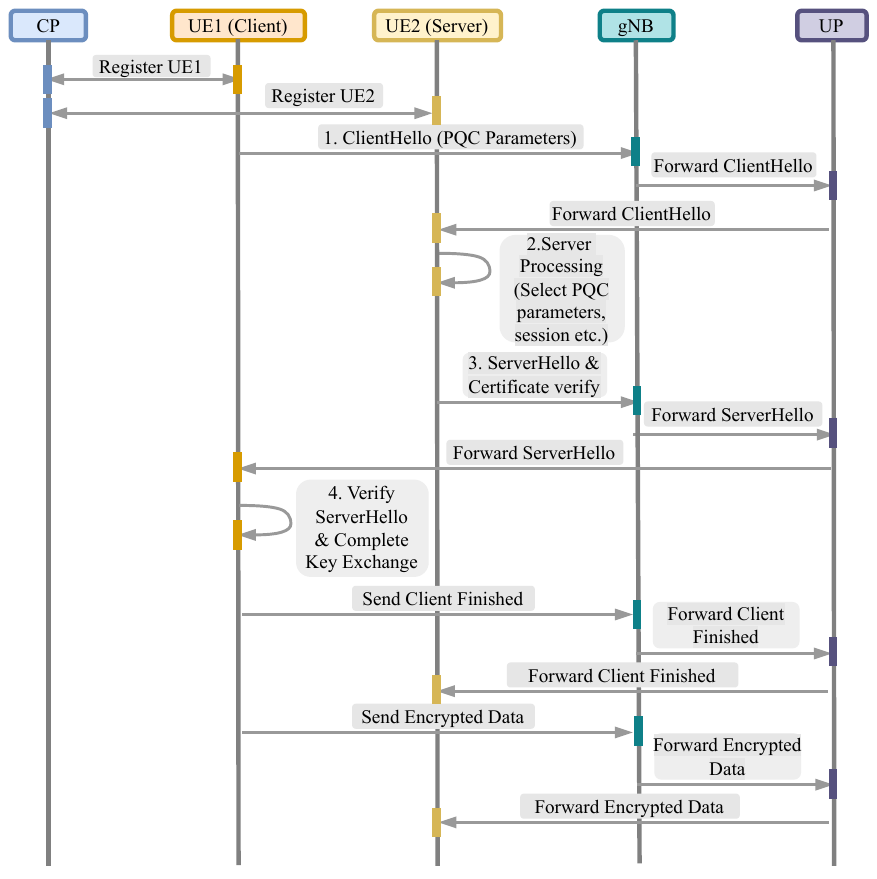}
    \caption{Sequence Diagram of TLS Handshake Process 
    }
    \label{fig:seq}    
    \vspace{-0.1in}
\end{figure}

\begin{itemize}
    \item \textit{Client Hello:} The client initiates the handshake by sending a ClientHello message. This message contains the supported PQC algorithms, cipher suites, and other cryptographic extensions. The message signals the start of secure session negotiation.
    \item \textit{Server Processing:} The server processes the ClientHello, verifies the parameters proposed by the client, and selects the cryptographic parameters (e.g., TLS\_AES\_128\_GCM\_SHA256, hqc128). The server responds with a ServerHello, specifies the selected algorithms, and sends its certificate and verification data.
    \item \textit{Certificate and Key Exchange:} The server sends its digital certificate to authenticate its identity, along with a signed message that verifies the handshake parameters. The client then verifies the server’s certificate to ensure its authenticity and correctness.
    \item \textit{Session Key Generation:} The client and the server jointly generate a session key using the selected key exchange algorithm (e.g., hqc128). This key is used to encrypt all subsequent data exchanged between the UEs.
    \item \textit{Handshake Completion:} The client sends a final message in which it confirms that the session key has been successfully generated and the handshake has been completed. The server acknowledges this message and thus officially establishes a secure connection.
\end{itemize}

Both client and server continuously monitor the connection to detect interruptions, packet loss, or retransmissions. System metrics such as CPU utilization, memory usage, bandwidth usage, and latency are logged throughout the data exchange. This data is crucial for evaluating the computational overhead associated with different PQC algorithms. 

\textbf{\textit{Data Exchange:}} Once a secure TLS session has been established, the data transfer between the UE client and the server is initiated. The client initiates the data transfer to the server via the established secure connection. The data packets are encrypted with the cipher suite selected during the handshake. 

\textbf{\textit{Performance Analysis:}}
The data logs of the individual VMs are aggregated to evaluate the impact of the PQC algorithms on network performance. Key metrics such as handshake latency, retransmission rates, and resource utilization are analyzed to identify computational bottlenecks and cryptographic overhead. 

\section{Experimental Evaluations} 

\subsection{Experimental Setup}
The experimental setup is structured to evaluate the impact of PQC algorithms on secure communication between UEs in a simulated 5G network. The experiment is conducted in a virtualized environment, where the host computer is equipped with a single-core Intel64 processor with 1700 MHz and 13,880 MB of available physical memory. The network interfaces are managed using VMware’s virtual Ethernet adapters, which set up isolated communication channels for data routing and monitoring. Five VMs are used to clearly separate the control and user layers and ensure different monitoring points for data collection. VM1 is referred to as the CP and manages authentication and session control, while VM2 acts as the UP and is responsible for data forwarding.  gNodeB is implemented in VM3 and facilitates the operation of the access network, while UE1 (VM4) and UE2 (VM5) handle secure communication, including the TLS handshake and data exchange processes. For consistency, each VM is assigned a static IP within the 192.168.234.0/24 subnet to ensure consistent data routing and monitoring. Data packets initiated by UE1 are routed through gNodeB to the UPF where they are forwarded to UE2 to simulate the data path. The UE VMs are configured to capture data packets at key interfaces, providing a comprehensive view of data flow, handshake latency, and cryptographic processing overhead. Open5GS and UERANSIM are used for the 5G core and access network simulation \cite{hoque2024post}. 

The security layer in each UE uses BoringSSL, configured for TLS 1.3 with support for PQC algorithms by liboqs. Quantum-resistant algorithms are used for key exchange and digital signatures, including ML-KEM, HQC, Falcon, and ML-DSA variants. Cryptographic events, including handshake initiation, key exchange completion, and data encryption/decryption, are logged. Data collection and monitoring are automated using structured scripts to ensure the consistency and reproducibility of the experimental results and shared with the community as an open-source as a Github page \cite{sanzi}. A comprehensive shell script has been developed to initiate multiple handshake attempts, log handshake timestamps, capture resource usage, and manage data extraction processes. The script takes input for the number of handshake attempts to perform per cryptographic configuration and records handshake initiation and termination timestamps. GNU Parallel \cite{tange2018gnu} is used to simulate simultaneous client connections and thus test scalability under cryptographic load. Concurrent client tests are performed with 10 and 20 clients to evaluate the impact of increasing cryptographic processing requirements on system resources. 

Network packet capture is performed using TShark \cite{tsoukalos2015using}, configured to monitor handshake packets, retransmission events, and data exchange traffic. Network traffic is logged to packet capture files (i.e., pcap files) for post-experiment analysis, with additional processing for identifying retransmission occurrences and data throughput rates. CPU and memory usage data are concurrently monitored using the \textit{ps aux} command, capturing resource consumption throughout the handshake and data exchange phases. This resource data is logged to a separate file to facilitate comparative analysis of cryptographic overhead across classical and PQC configurations.

Data analysis is conducted using Python and batch scripts, focusing on calculating average handshake latency, total bandwidth, CPU and memory utilization, and retransmission rates. The analysis script processes logs to extract handshake timestamps, calculate latency for each handshake attempt, and determine average values over 50 iterations. The CPU and memory usage data are averaged to evaluate the computational overhead associated with the PQC operations. Visualization is done with the Python library matplotlib. The experimental scenarios are designed to measure the baseline performance with classical algorithms and evaluate the PQC overhead with different combinations of KEMs and digital signatures. These scenarios provide a comprehensive comparison of computational costs and network performance with different cryptographic loads.


\subsection{Experimental Metrics}
\textit{Max CPU Usage (\%):} This metric captures the peak CPU utilization percentage observed on the server side during the TLS data transmission phase
. It reflects the computational overhead caused by cryptographic operations (e.g., key encapsulation, decapsulation, signing, and verification), and memory handling.

\textit{Latency (ms):} The average time, measured in milliseconds, required to complete a full TLS handshake between the client and server. This metric reflects how quickly a secure session can be established.

\textit{Bandwidth (KB/s):}  The total throughput used during the secure communication session, including both transmission (TX) and reception (RX), reported in kilobytes per second. 


\textit{Retransmission (Retx) Rate (\%):} The retransmission rate is the percentage of packets that are retransmitted in a communication session relative to the total number of packets sent. It is calculated as:
\[
\text{Retx Rate (\%)} = \left( \frac{\text{Number of Retransmitted Packets}}{\text{Total Number of Packets Sent}} \right) \times 100
\]

A higher retransmission rate typically indicates network congestion, packet loss, or communication errors that necessitate packet retransmission to maintain data integrity.

\begin{table*}[h!]
\centering
\scriptsize
\caption{Post Quantum + Classic KEM with Falcon Signature (Variants separated by lines)}
\label{Tab:falcon}
\begin{tabular}{lcccc}
\toprule
\textbf{KEM\_SIGNATURE} & \textbf{Max CPU Usage (\%)} & \textbf{TLS Handshake Latency (ms)} & \textbf{Bandwidth (KB/s)} & \textbf{Retransmission Rate (\%)} \\
\midrule
X25519\_falcon512 & 0.50 & 35 & 15.22 & 0.5739
 \\
secp384r1\_falcon512 & 1.20 & 34 & 16.60 & 0.2899
 \\
secp521r1\_falcon512 & 1.40 & 41 & 16.66 & 0.6258
 \\
mlkem512\_falcon512 & 0.40 & 35 & 21.75 & 0.9988
 \\
mlkem768\_falcon512 & 0.50 & 36 & 25.30 & 1.2115
 \\
mlkem1024\_falcon512 & 1.00 & 42 & 28.02 & 1.4866
 \\

hqc128\_falcon512 & 1.40 & 42 & 43.84 & 1.1551
 \\
hqc192\_falcon512 & 2.50 & 66 & 68.09 & 1.0508\\
hqc256\_falcon512 & 4.80 & 91 & 92.44 & 0.3913
 \\
\midrule
X25519\_falcon1024 & 0.70 & 32 & 25.34 & 1.2155
 \\
secp384r1\_falcon1024 & 1.20 & 31 & 26.98 & 0.6689
 \\
secp521r1\_falcon1024 & 2.00 & 41 & 26.07 & 1.3215
 \\
mlkem512\_falcon1024 & 0.60 & 32 & 33.28 & 0.9872
 \\
mlkem768\_falcon1024 & 0.60 & 33 & 35.01 & 0.7279
 \\
mlkem1024\_falcon1024 & 0.70 & 37 & 38.60 & 1.3429
 \\

hqc128\_falcon1024 & 1.10 & 42 & 53.47 & 1.5151
 \\
hqc192\_falcon1024 & 3.70 & 67 & 76.12 & 0.4742
 \\
hqc256\_falcon1024 & 4.50 & 85 & 100.11 & 0.3215
 \\
\bottomrule
\end{tabular}
\end{table*}

\begin{table*}[h!]
\centering
\scriptsize
\caption{Post Quantum + Classic KEM with SPHINCS+ Signature (Variants separated by lines)}
\label{Tab:sphinc}
\begin{tabular}{lcccc}
\toprule
\textbf{KEM\_SIGNATURE} & \textbf{Max CPU Usage (\%)} & \textbf{TLS Handshake Latency (ms)} & \textbf{Bandwidth (KB/s)} & \textbf{Retransmission Rate (\%)} \\
\midrule
X25519\_sphincssha2128f & 3.40 & 61 & 141.83 & 0.1722
 \\
secp384r1\_sphincssha2128f & 4.10 & 62 & 142.56 & 0.2063
 \\
secp521r1\_sphincssha2128f & 4.10 & 66 & 143.21 & 0.0692
 \\
mlkem512\_sphincssha2128f & 2.70 & 56 & 149.95 & 0.0335
 \\
mlkem768\_sphincssha2128f & 3.10 & 65 & 148.17 & 0.1620
 \\
mlkem1024\_sphincssha2128f & 3.70 & 61 & 153.92 & 0.2583
 \\

hqc128\_sphincssha2128f & 4.50 & 64 & 165.83 & 0.2183
 \\
hqc192\_sphincssha2128f & 4.70 & 77 & 182.98 & 0.0836
 \\
hqc256\_sphincssha2128f & 6.80 & 102 & 196.41 & 0.4827
 \\
\midrule
X25519\_sphincssha2192f & 4.90 & 84 & 265.38 & 0.0385
 \\
secp384r1\_sphincssha2192f & 6.80 & 90 & 261.66 & 0.1694
 \\
secp521r1\_sphincssha2192f & 5.50 & 88 & 263.02 & 0.1331
 \\
mlkem512\_sphincssha2192f & 4.80 & 95 & 260.85 & 0.0382
 \\
mlkem768\_sphincssha2192f & 4.80 & 94 & 263.55 & 0.0934
 \\
mlkem1024\_sphincssha2192f & 5.50 & 96 & 265.45 & 0.0545
 \\

hqc128\_sphincssha2192f & 4.90 & 79 & 302.87 & 0.2740
 \\
hqc192\_sphincssha2192f & 6.20 & 92 & 304.16 & 0.1267
 \\
hqc256\_sphincssha2192f & 8.70 & 116 & 309.22 & 0.1199
 \\
\midrule
X25519\_sphincssha2256f & 6.90 & 100 & 359.25 & 0.3590
 \\
secp384r1\_sphincssha2256f & 7.80 & 113 & 337.42 & 0.1934
 \\
secp521r1\_sphincssha2256f & 9.10 & 112 & 337.88 & 0.3019
 \\
mlkem512\_sphincssha2256f & 7.80 & 94 & 363.26 & 0.3381
 \\
mlkem768\_sphincssha2256f & 8.20 & 97 & 363.24 & 0.3150
 \\
mlkem1024\_sphincssha2256f & 8.70 & 125 & 333.38 & 0.1737
 \\

hqc128\_sphincssha2256f & 9.30 & 112 & 359.40 & 0.2948
 \\
hqc192\_sphincssha2256f & 6.30 & 119 & 373.70 & 0.2292
 \\
hqc256\_sphincssha2256f & 10.60 & 140 & 374.01 & 0.0981
 \\
\bottomrule
\end{tabular}
\end{table*}

\begin{table*}[h!]
\centering
\scriptsize
\caption{Post Quantum + Classic KEM with ML-DSA Signature (Variants separated by lines)}
\label{Tab:dilithium}
\begin{tabular}{lcccc}
\toprule
\textbf{KEM\_SIGNATURE} & \textbf{Max CPU Usage (\%)} & \textbf{TLS Handshake Latency (ms)} & \textbf{Bandwidth (KB/s)} & \textbf{Retransmission Rate (\%)} \\
\midrule
X25519\_mldsa44 & 0.40 & 23 & 34.20 & 0 \\
secp384r1\_mldsa44 & 1.20 & 31 & 34.72 & 0 \\
secp521r1\_mldsa44 & 1.70 & 32 & 34.56 & 0 \\
mlkem512\_mldsa44 & 0.20 & 23 & 41.14 & 0.0921
 \\
mlkem768\_mldsa44 & 0.20 & 24 & 44.72 & 0.1689

 \\
mlkem1024\_mldsa44 & 0.30 & 24 & 48.84 & 0 \\

hqc128\_mldsa44 & 1.60 & 40 & 60.82 & 0.0673
 \\
hqc192\_mldsa44 & 2.60 & 57 & 84.56 & 0 \\
hqc256\_mldsa44 & 4.20 & 80 & 108.53 & 0.0387
 \\
\midrule
X25519\_mldsa65 & 0.40 & 25 & 46.30 & 0 \\
secp384r1\_mldsa65 & 1.20 & 31 & 47.77 & 0.1681
 \\
secp521r1\_mldsa65 & 1.70 & 39 & 44.13 & 0 \\
mlkem512\_mldsa65 & 0.30 & 27 & 52.91 & 0.2333
 \\
mlkem768\_mldsa65 & 0.30 & 25 & 56.63 & 0 \\
mlkem1024\_mldsa65 & 0.30 & 28 & 59.45 & 0 \\

hqc128\_mldsa65 & 1.50 & 39 & 71.24 & 0.1258
 \\
hqc192\_mldsa65 & 3.00 & 54 & 95.62 & 0.0456
 \\
hqc256\_mldsa65 & 4.80 & 83 & 116.12 & 0.0360
 \\
\midrule
X25519\_mldsa87 & 0.40 & 30 & 62.22 & 0 \\
secp384r1\_mldsa87 & 1.10 & 33 & 60.54 & 0.1343
 \\
secp521r1\_mldsa87 & 1.50 & 33 & 60.56 & 0.2015
 \\
mlkem512\_mldsa87 & 0.30 & 26 & 68.25 & 0.0674
 \\
mlkem768\_mldsa87 & 0.30 & 29 & 72.56 & 0.0629
 \\
mlkem1024\_mldsa87 & 0.30 & 29 & 76.50 & 0 \\

hqc128\_mldsa87 & 1.60 & 39 & 85.64 & 0 \\
hqc192\_mldsa87 & 3.20 & 58 & 107.54 & 0 \\
hqc256\_mldsa87 & 3.30 & 81 & 129.13 & 0.0336
 \\
\bottomrule
\end{tabular}
\end{table*}

\mycomment{
\begin{table*}[htbp]
\centering
\caption{Categorized Performance of PQC KEM+Signature Combinations in UE-to-UE Communication}
\label{tab:pqc_performance}
\begin{tabular}{|l|c|c|c|c|c|}
\hline
\textbf{KEM + Signature} & \textbf{Latency (ms)} & \textbf{Bandwidth (KB/s)} & \textbf{Avg Time (s)} & \textbf{Total Packets} & \textbf{Retransmission Rate (\%)} \\
\hline
\multicolumn{6}{|c|}{\textbf{Efficient (High-Performance PQC)}} \\
\hline
ML-KEM768 + Dilithium5 & 27 & 627.94 & 0.00150 & 2546 & 0.28 \\
ML-KEM512 + Dilithium5 & 26 & 524.67 & 0.00119 & 2389 & 0.38 \\
ML-KEM1024 + Dilithium5 & 32 & 487.03 & 0.00137 & 2704 & 0.07 \\
X25519 + Falcon512       & 23 & 139.50 & 0.00167 & 1115 & 0.27 \\
\hline
\multicolumn{6}{|c|}{\textbf{Moderate (Hybrid / Classical ECC-Based)}} \\
\hline
secp384r1 + Falcon512     & 30 & 116.18 & 0.00184 & 1122 & 0.45 \\
secp521r1 + Falcon512     & 37 & 105.80 & 0.00169 & 1275 & 0.39 \\
secp384r1 + Dilithium2    & 29 & 266.60 & 0.00150 & 1586 & 0.06 \\
X25519 + Dilithium2       & 27 & 261.89 & 0.00160 & 1597 & 0.38 \\
\hline
\multicolumn{6}{|c|}{\textbf{Inefficient (SPHINCS+, High-Overhead)}} \\
\hline
ML-KEM1024 + SPHINCS\texttt{+}-SHA2192f & 138 & 579.39 & 0.00206 & 9265 & 3.50 \\
secp521r1 + SPHINCS\texttt{+}-SHA2128f  & 178 & 223.01 & 0.00541 & 5122 & 4.59 \\
secp384r1 + SPHINCS\texttt{+}-SHA2128f  & 143 & 268.22 & 0.00446 & 5146 & 4.78 \\
ML-KEM512 + SPHINCS\texttt{+}-SHA2128f  & 72  & 540.55 & 0.00215 & 5139 & 2.65 \\
\hline
\end{tabular}
\end{table*}
}

\subsection{Performance Analysis}

Tables \ref{Tab:falcon}, \ref{Tab:sphinc}, and \ref{Tab:dilithium} provide the performance outcomes for Falcon, SPHINCS+, and ML-DSA, respectively, using both classical and post-quantum Key Encapsulation Mechanisms (KEM).

The \textit{ML-DSA} variants consistently demonstrate the lowest CPU usage across all evaluated categories, with maximum utilization levels starting from 0.20\%. 
This efficiency is most pronounced in the mldsa44 configuration, where the average TLS handshake latency is just 23 ms, underscoring its suitability for latency-sensitive applications in 5G networks. The retransmission rate is also notably low, with no or only 1-2 packets retransmitted per session, indicating a reliable transmission profile.


In contrast, the \textit{SphincsSHA} variants, particularly the sphincssha2256f configuration, impose substantial computational overhead, with maximum CPU utilization peaking at 10.60\% for HQC256\_sphincssha2256f. The average TLS handshake latency reaches 140 ms, more than three times that of MLDSA, indicating a significant processing burden. Bandwidth consumption is similarly elevated, averaging 260 KB/s, reflecting the larger key sizes and signature lengths associated with SphincsSHA. 

These findings reveal that SphincsSHA, despite its strong security guarantees, presents notable performance drawbacks. The elevated CPU usage and prolonged handshake latency render it less suitable for latency-sensitive applications in 5G. For instance, the required expected end-to-end latency for advanced 5G applications such as specialist/surgeons training, indoor and localized outdoor navigation, AR-based driver training, Flight pilot training, is less than 20ms, whereas the minimum latency in SPHINCS+ variant combination shows 56ms, making it not suitable for advanced 5G applications \cite{siriwardhana2021survey}. The cloud-based mobile augmented reality applications' expected latency is higher as 50ms, yet lower than the minimum latency of this variant. However, for secure archival and digital signatures in data centers, SphincsSHA’s conservative security posture remains advantageous despite its operational overhead \cite{bernstein2015sphincs}.

\textit{Falcon}-based combinations offer a balanced performance profile, characterized by moderate CPU usage and consistent bandwidth consumption. The average CPU utilization across Falcon variants is 1.60\%, with the lowest value observed in mlkem512\_falcon512 at 0.40\%, demonstrating efficient key exchange processes. The average handshake latency remains moderate at 45 ms across combinations, aligning closely with that of MLDSA and notably lower than the SphincsSHA variants. Bandwidth usage is relatively stable across both Falcon512 and Falcon1024 configurations, suggesting efficient data transmission management.
However, while Falcon exhibits balanced metrics, the packet loss rates in higher-security configurations, such as hqc128\_falcon1024, rise to 1.5\%, indicating potential congestion issues under increased data loads. 

Overall, \textit{\textbf{for lightweight, latency-sensitive applications}}, ML-DSA variants, particularly mldsa44 and mldsa65, offer the best trade-off between minimal computational overhead and secure data exchange. These configurations maintain low packet counts and minimal retransmissions, making them well-suited for real-time communication scenarios where low computational overhead is essential. Notably, the mlkem512\_mldsa65 configuration exhibits higher retransmissions, suggesting potential buffer management inefficiencies that could be mitigated through optimization strategies.
\textit{\textbf{For security-intensive environments}}, SphincsSHA configurations, especially sphincssha2256f, provide robust cryptographic integrity but at the expense of significantly higher CPU usage, latency, and bandwidth consumption. The high retransmission rates indicate vulnerability to data congestion, making SphincsSHA more appropriate for secure archival, data integrity verification, and long-term data storage rather than real-time UE-to-UE communications.
\textit{\textbf{For balanced performance across security levels}}, Falcon-based combinations achieve a stable performance profile with moderate latency and bandwidth usage. The consistent CPU utilization in X25519\_falcon512 and secp384r1\_falcon1024 suggests their suitability for general-purpose secure communication. However, the increased retransmission rates in higher-security variants, such as mlkem1024\_falcon512 and mlkem1024\_falcon1024, suggest potential data congestion under high-throughput conditions.

\begin{figure}
    \centering
    \includegraphics[width=.9\linewidth]{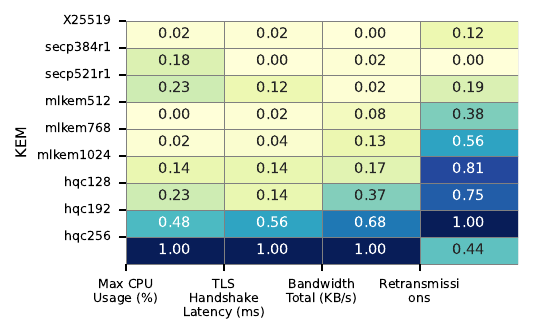}
    \caption{Heatmap of Normalized Performance Metrics of Falcon512 Across KEMs }
    \vspace{-0.1in}
     \label{fig:kem-a}
\end{figure}

\begin{figure}
    \centering
    \includegraphics[width=.9\linewidth]{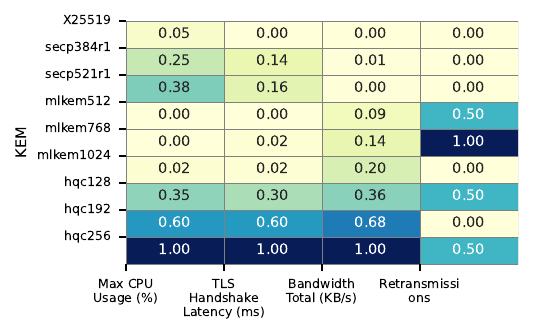}
    \caption{Heatmap of Normalized Performance Metrics of MLDSA44 Across KEMs }
    \vspace{-0.1in}
     \label{fig:kem-b}
\end{figure}

\begin{figure}
    \centering
    \includegraphics[width=0.9\linewidth]{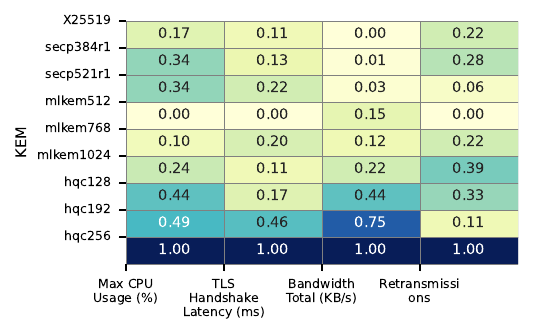}
    \caption{Heatmap of Normalized Performance Metrics of Sphincssha2128fsimple Across KEMs}
    \vspace{-0.1in}
    \label{fig:kem-c}
\end{figure}

\begin{figure}
    \centering
    \includegraphics[width=.9\linewidth]{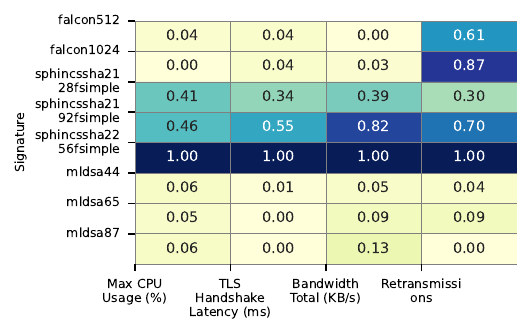}
    \caption{Heatmap of Normalized Performance Metrics of HQC128 Across Signatures}
    \vspace{-0.1in}
   \label{fig:sign-a}
\end{figure}

\begin{figure}
    \centering
    \includegraphics[width=.9\linewidth]{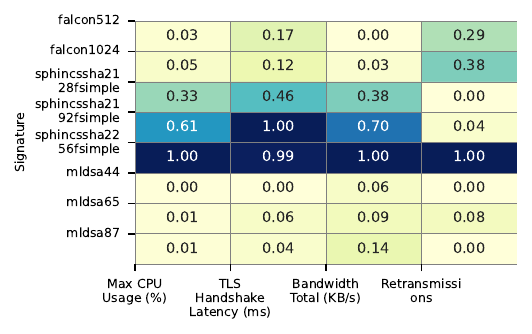}
    \caption{Heatmap of Normalized Performance Metrics of MLKEM512 Across Signatures }
    \vspace{-0.1in}
    \label{fig:sign-b}
\end{figure}

\mycomment{
\begin{figure}[htbp]
    \centering
    \begin{minipage}{0.5\textwidth}
        \centering
        \includegraphics[width=\linewidth]{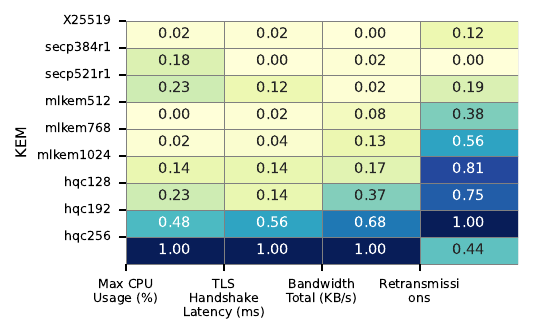}
        \subcaption{Figure 1}\label{fig:1}
    \end{minipage}%
    \hfill
    \begin{minipage}{0.5\textwidth}
        \centering
        \includegraphics[width=\linewidth]{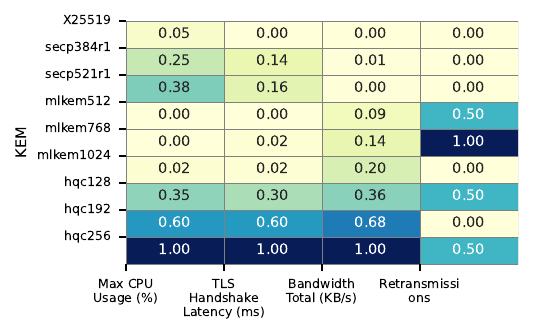}
        \subcaption{Figure 2}\label{fig:2}
    \end{minipage}%
    \hfill
    \begin{minipage}{0.5\textwidth}
        \centering
        \includegraphics[width=\linewidth]{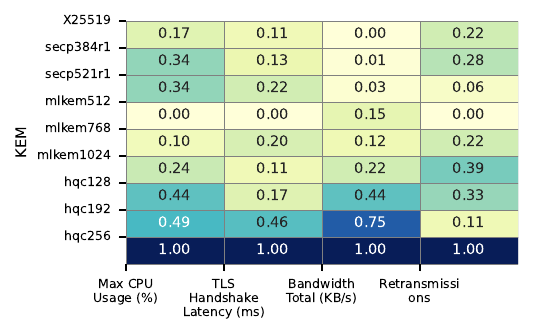}
        \subcaption{Figure 3}\label{fig:3}
    \end{minipage}
    
    \vspace{0.5cm}
    
    \begin{minipage}{0.5\textwidth}
        \centering
        \includegraphics[width=\linewidth]{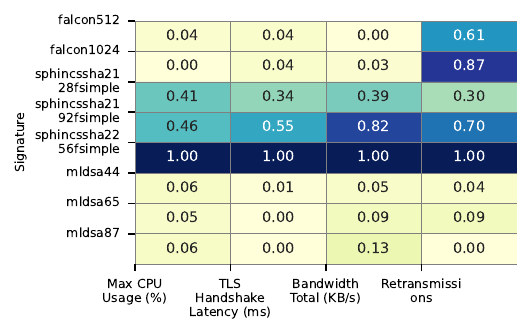}
        \subcaption{Figure 4}\label{fig:4}
    \end{minipage}%
    \hfill
    \begin{minipage}{0.5\textwidth}
        \centering
        \includegraphics[width=\linewidth]{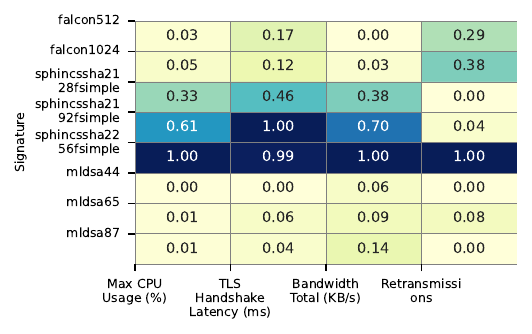}
        \subcaption{Figure 5}\label{fig:5}
    \end{minipage}%
    \hfill
    \caption{abc}
    \label{abc}
\end{figure}
}

Moreover, the comparative heatmaps for Falcon512 (Fig. \ref{fig:kem-a}), MLDSA44 (Fig. \ref{fig:kem-b}), and SphincsSHA2128fsimple (Fig. \ref{fig:kem-c}) provide visualizations of the normalized performance metrics across multiple KEM variants.
In these heatmaps, for a given metric value \( x \), the normalized value \( x_{\text{norm}} \) is computed as: \[
x_{\text{norm}} = \frac{x - \min(X)}{\max(X) - \min(X)}
\]
where \( X \) denotes the set of all values for that metric across all configurations; \( \min(X) \) and \( \max(X) \) are the minimum and maximum values of the metric, respectively.

In Fig. \ref{fig:kem-a}, Falcon512 shows optimal performance with low CPU usage (0.02 for X25519, 0.23 for mlkem512) and minimal latency (0.02 for X25519), but HQC256 peaks at 1.00 in CPU, latency, and bandwidth, indicating heavy resource demand. MLDSA44 maintains low overhead with X25519 and mlkem512 both at 0.02 in CPU and latency, but HQC256 again spikes to 1.00 in all metrics, while mlkem1024 experiences high retransmissions (1.00). SphincsSHA2128fsimple distributes CPU, latency, and bandwidth more evenly, with X25519 and mlkem512 at 0.17 and 0.00 in CPU, but HQC256 remains the most intensive (1.00 across metrics), indicating its focus on security over efficiency.

Similarly, Fig. \ref{fig:sign-a} and  \ref{fig:sign-b} present the heatmap of normalized performance metrics of Level-1 PQC Signatures HQC and MLKEM, respectively. HQC128  shows minimal CPU usage (0.04) and latency (0.04) for Falcon512, while SphincsSHA2256fsimple peaks across all metrics (1.00), indicating intensive processing and bandwidth demand. MLKEM512 maintains lower CPU and retransmissions, but SphincsSHA2256fsimple exhibits the highest resource usage (1.00 CPU, 0.99 latency, 1.00 bandwidth), highlighting its computational overhead. Overall, falcon1024 records substantial retransmissions (0.87) under HQC128, suggesting instability, whereas MLKEM512 manages more balanced performance despite increasing signature complexity.

\begin{table*}[h!]
\centering
\scriptsize
\caption{Categorized Performance of PQC KEM+Signature Combinations in UE-to-UE Communication}
\label{tab:pqc_performance}
\begin{tabular}{|l|c|c|c|c|}
\hline
\textbf{KEM + Signature} & \textbf{Latency (ms)} & \textbf{Bandwidth (KB/s)} & \textbf{Total Packets} & \textbf{Number of Retransmissions} \\
\hline
\multicolumn{5}{|c|}{\textbf{Efficient (High-Performance PQC)}}\\
\hline
X25519\_mldsa44 & 23 & 34.20 & 985 & 0 \\ \hline 
mlkem512\_mldsa44 & 23 & 41.14 & 1086 & 1 \\ \hline 
secp384r1\_sphincssha2128f & 62 & 142.56 & 2908 & 6 \\
\hline
\multicolumn{5}{|c|}{\textbf{Moderate (Balanced Performance)}}\\
\hline
mlkem768\_mldsa65 & 25 & 56.63 & 1386 & 0 \\ \hline 
secp521r1\_mldsa87 & 33 & 60.56 & 1489 & 3 \\ \hline 
secp521r1\_sphincssha2256f & 112 & 337.88 & 7287 & 22 \\
\hline
\multicolumn{5}{|c|}{\textbf{High Security (Resource-Intensive)}}\\
\hline
HQC256\_sphincssha2256f & 140 & 374.01 & 8159 & 8 \\ \hline 
HQC256\_mldsa87 & 81 & 129.13 & 2978 & 1 \\
mlkem1024\_sphincssha2256f & 125 & 333.38 & 7485 & 13 \\
\hline 
\end{tabular}
\end{table*}

Table \ref{tab:pqc_performance} provides a summary of the performance of the PQC combinations. The performance results of our UE-to-UE experiments using PQC algorithms are mostly in line with theoretical expectations based on the cryptographic design principles \cite{paquin2020benchmarking}, although some results deserve a closer examination. As expected, the combinations with ML-KEM and ML-DSA - both lattice-based and selected by NIST for standardization - shows excellent performance on all key metrics.  X25519\_MLDSA44 and MLKEM512\_MLDSA44 are optimal for real-time communications, providing minimal latency and retransmissions. For applications demanding a balance between security and performance, MLKEM768\_MLDSA65 and secp521r1\_MLDSA87 offer moderate latency and secure data exchange, albeit with occasional retransmissions. For scenarios prioritizing cryptographic robustness, HQC256\_Sphincssha2256f and MLKEM1024\_Sphincssha2256f provide the highest security levels but require careful bandwidth management and retransmission mitigation to prevent data congestion and ensure reliable communication. These results reinforce the view that lattice-based techniques are not only secure but also well suited to real-time applications such as UE-to-UE communications in 5G environments, where both speed and reliability are critical.

\begin{table*}[h!]
    \centering
\scriptsize
    \caption{Performance Metrics for Scalability Test}
    \begin{tabular}{lcccccc}
        \toprule
        Algorithm & Clients & Max CPU Usage (\%) & Avg Handshake Latency (ms) & Bandwidth (KB/s) & Retransmission Rate (\%) \\
        \midrule
        mlkem512\_mldsa44 & 10 & 2.70 & 48.2 & 366.973 & 0.1726 \\
                         & 20 & 3.70 & 178.8 & 460.885 & 0.2157 \\
        \midrule
        hqc128\_mldsa44 & 10 & 13.30 & 86 & 499.767 & 0.1808 \\
                        & 20 & 19.30 & 369.55 & 493.0865 & 0.1045 \\
        \midrule
        mlkem512\_falcon512 & 10 & 6.00 & 62.7 & 178.35 & 0.1546 \\
                            & 20 & 8.20 & 245.75 & 204.0255 & 0.3547 \\
        \midrule
        hqc128\_falcon512 & 10 & 5.20 & 51.6 & 408.821 & 0.1423 \\
                          & 20 & 8.40 & 343.2 & 375.061 & 0.2572 \\
        \midrule
        mlkem512\_sphincssha2128f & 10 & 10.30 & 210.9 & 918.666 & 0.0185 \\
                                  & 20 & 22.30 & 620.8 & 919.1485 & 0.0217 \\
        \midrule
        hqc128\_sphincssha2128f & 10 & 21.90 & 297.1 & 865.1565 & 0.0368 \\
                                & 20 & 31.30 & 882.55 & 786.774 & 0.0412 \\
        \bottomrule
    \end{tabular}
\end{table*}

\subsubsection{\textbf{Scalability Stress Test}}
The performance analysis focuses on six PQC algorithm combinations, all operating at Level 1 security level, are also evaluated under two different scenarios for scalability: 10 clients and 20 clients. The algorithms include mlkem512\_mldsa44, hqc128\_mldsa44, mlkem512\_falcon512, hqc128\_falcon512, mlkem512\_sphincssha2128f, and hqc128\_sphincssha2128f. The analysis assesses four key performance metrics: CPU usage, latency, bandwidth, and retransmission rates.

\textbf{CPU usage:} CPU usage increases consistently with the number of clients across all algorithm combinations, indicating the computational cost of managing higher connection loads. The highest CPU usage is observed in hqc128\_sphincssha2128f, rising from 21.9\% (10 clients) to 31.3\% (20 clients). This combination involves the computationally intensive sphincssha2128f signature scheme, known for its large signature sizes and complex hash operations. Conversely, the lowest CPU usage is recorded for mlkem512\_mldsa44, with values of 2.7\% and 3.7\%, suggesting that the mldsa44 signature scheme is relatively lightweight in processing compared to other combinations. 

\textbf{Latency:}
Latency metrics indicate significant variability across the algorithm combinations. The most substantial increase in latency is observed in mlkem512\_sphincssha2128f, which rises from 210.9 ms to 620.8 ms, reflecting the computational overhead associated with the sphincssha2128f signature scheme.
In contrast, mlkem512\_falcon512 demonstrates a moderate latency increase from 62.7 ms to 245.75 ms, suggesting that falcon512 maintains more consistent performance under higher client loads. This pattern indicates that falcon512, a lattice-based signature scheme, may provide a more balanced trade-off between cryptographic complexity and processing time.

\textbf{Bandwidth usage:} Bandwidth usage generally increases as the number of clients rises, with notable differences in algorithm combinations. For instance, hqc128\_mldsa44 exhibits a substantial increase in bandwidth usage from 86 KB/s to 369.55 KB/s, indicating higher data transfer requirements as client connections increase.
Interestingly, mlkem512\_sphincssha2128f maintains relatively stable bandwidth usage, increasing marginally from 918.666 KB/s to 919.1485 KB/s, suggesting efficient data handling despite increased computational processing. This stability may indicate that the sphincssha2128f signature scheme, despite its higher CPU and latency costs, manages data transfer more effectively under load. However, it costs more in terms of transmission times (i.e., latency).

\textbf{Retransmission rates:}
Retransmission rates generally increase with higher client loads, reflecting potential congestion and packet loss. The most pronounced increase is observed in mlkem512\_falcon512, where the retransmission rate rises from 0.1546\% to 0.3547\%, suggesting that falcon512 may be more susceptible to packet loss under heavier client loads. Similarly, hqc128\_falcon512 demonstrates a substantial increase in retransmission rates from 0.1423\% to 0.2572\%, suggesting potential data congestion challenges associated with Falcon512 under increased traffic conditions. In contrast, mlkem512\_sphincssha2128f maintains the lowest retransmission rates at 0.0185\% (10 clients) and 0.0217\% (20 clients), indicating robust data transmission stability despite its higher latency. This stability is likely attributable to SphincsSHA2128f’s hash-based structure, which appears more resistant to packet fragmentation under intensive data transmission. A notable trend is observed in hqc128\_mldsa44, where the retransmission rate decreases from 0.1808\% to 0.1045\% despite increased client loads. While HQC’s error-correcting capabilities may contribute to this reduction, other system-level factors could also be influencing this outcome. 

Based on the observed performance profiles, mlkem512\_mldsa44 and hqc128\_falcon512 exhibit balanced latency and CPU usage, making them suitable for mainstream 5G applications \cite{siriwardhana2021survey}. However, for ultra-low latency and high thoughput scenarios, mlkem512\_sphincssha2128f and hqc128\_sphincssha2128f show strong throughput but face scalability challenges. With targeted optimizations, these algorithms hold significant potential to excel in data-intensive, bandwidth-heavy 5G and emerging 6G deployment.



\section{Conclusion}
As the security foundations of classical cryptographic systems become increasingly vulnerable to quantum computing, it is critical to evaluate the feasibility of deploying PQC in real-world wireless environments. This paper presented a comprehensive experimental analysis of NIST-selected PQC algorithms integrated into UE-to-UE communication within a 5G network stack. Our results indicate that ML-KEM combined with ML-DSA offers a highly efficient solution for secure communication, with low latency, minimal CPU overhead, and low retransmission rates, making it ideal for latency-sensitive mobile applications. In contrast, SPHINCS+ and HQC combinations introduce significant processing and transmission overhead, suggesting their suitability for scenarios where long-term security outweighs real-time performance, such as secure archival or high-assurance data exchange. Scalability tests further demonstrated that while some PQC schemes maintain stable performance under increased client loads, others suffer from increased latency and CPU usage, underscoring the importance of cryptographic agility in system design. Overall, this work highlights the practical trade-offs in integrating PQC into 5G systems and provides data-driven insights to guide future secure protocol design for 5G, 6G, and beyond. Future work will explore scalable implementation strategies, real-world deployment challenges, and the integration of PQC with evolving 6G security architectures.

\balance

\bibliographystyle{IEEEtran}
\bibliography{References}

\end{document}